\documentstyle[aas2pp4]{article}

\def\msun{{\,M_\odot}}

\def\refindent{\par\noindent\hangindent=3pc\hangafter=1 }
\def\aa#1#2#3{\refindent#1, A\&A, #2, #3}
\def\aasup#1#2#3{\refindent#1, A\&AS, #2, #3}

\def\apj#1#2#3{\refindent#1, {\it ApJ}, {\bf#2}, #3.}
\def\apjlett#1#2#3{\refindent#1, {\it ApJL}, {\bf #2}, #3.}

\def\>{$>$}
\def\<{$<$}

\def\simlt{\lower.5ex\hbox{$\; \buildrel < \over \sim \;$}}
\def\simgt{\lower.5ex\hbox{$\; \buildrel > \over \sim \;$}}
\def\sqr#1#2{{\vcenter{\hrule height.#2pt
      \hbox{\vrule width.#2pt height#1pt \kern#1pt
         \vrule width.#2pt}
      \hrule height.#2pt}}}

\begin{document}

\centerline{Submitted to the Astrophysical Journal (Letters)}

\bigskip
\title{Magnetic Flares and State Transitions in Galactic Black Hole and \\
Neutron Star Systems}
\author{Sergei Nayakshin$^{\dagger}$ and Fulvio Melia$^*$
\thanks{Presidential Young Investigator.}}
\affil{$^{\dagger}$Physics Dept., The University of Arizona, 
Tucson AZ 85721\\
$^*$Physics Dept. \& Steward Observatory,
The University of Arizona, Tucson AZ 85721}




\altaffiltext{1}{Presidential Young Investigator.}


\begin{abstract}
We here examine the conditions of the two-phase disk model under which
magnetic flares arise above the cold accretion disk due to magnetic
buoyancy and produce X-rays via Comptonization of the disk's soft
radiation. We find that the disk's ability to produce strong magnetic 
flares is substantially diminished in its radiation dominated 
regions due to the diffusion of radiation into the magnetic flux tubes. 
Using a simplified, yet physically self-consistent, model that takes this 
effect into account, we show that the hard X-ray spectrum of some GBHCs can be
explained as the X-ray emission by magnetic flares only when the disk's
bolometric luminosity is a relatively small fraction ($\sim$ 5\%) of the
Eddington value, $L_{\rm Edd}$. Further, we compute the hard ($20-200$ keV) and
soft ($1-20$ keV) X-ray power as a function of the disk's luminosity,
and find an excellent agreement with the available data for GBHC
transient and persistent sources. We conclude that the observed
high-energy spectrum of stellar-sized accretion disk systems can be 
explained by Comptonization of the disk's soft radiation by 
the hot gas trapped inside the magnetic flares 
when the luminosity falls in the range $\sim 10^{-3}-10^{-1}\times L_{\rm Edd}$.
For higher luminosities, another emission mechanism must be at work. 
For lower luminosities, the X-ray emissivity may still be dominated
by magnetic flares, but this process is more likely to be
thermal or non-thermal bremstrahlung, so that the X-ray spectrum below
$\sim 10^{-3}L_{\rm Edd}$ may be quite distinct from the typical hard
spectrum for higher luminosities.
\end{abstract}




%

\section{Introduction}

Many Galactic Black hole Candidates (GBHC) display two relatively well
defined (``hard'' and ``soft'') spectral states (Grove et al. 1997; Zhang 
et al. 1997a).  There is no well established theory as to why these 
transitions occur, nor is there a complete understanding of what constitutes
either of these definite states, though a variety of models have been suggested. 

Recently, Nayakshin \& Melia (1997a) have shown that the spectrum of 
Cygnus X-1 in its hard state can be explained by the two-phase model 
with magnetic flares accounting for the patchy overlying corona, if one takes
into account the time-dependent nature of these flares (also referred to below 
as Active Regions [AR]). The two-phase model was originally applied
to Seyfert Galaxies by Haardt \& Maraschi (1991, 1993) (though the model of
a hot corona above the cold accretion disk was suggested long before that for
the Galactic source Cygnus X-1; see, e.g., Liang \& Price 1977) in order
to explain the observed multi-wave band 
spectrum of these sources, and appears to be successful in accounting
for many of the observed features in these objects (see also Galeev, Rosner 
\& Vaiana 1979; Field \& Rogers 1993; Haardt, Maraschi \& Ghisellini 1994; 
Svensson 1996; Nayakshin \& Melia 1997b,c,d). Nayakshin \& Melia (1997a) 
also showed that the differences in the spectrum, the ionization state of 
the cold disk, and the iron lines and edges in the hard state of Cygnus X-1 
compared to the corresponding observables in Seyfert Galaxies may be a natural 
consequence of a change in the physical state of the {\it transition layer}, 
where the hard X-rays from the ARs are reflected or reprocessed. Therefore, 
the magnetic flare-cold disk model appears to be a ``unifying'' link between
Seyferts and GBHCs in their hard state.

In view of the fact that GBHCs display the softer spectral state as well, it 
would be helpful to examine more fully the physics of magnetic flares in a 
cold disk to see if these can provide an explanation for both states, and the 
transitions between them. In this {\it Letter}, we attempt to determine whether 
or not the flare activity is correlated with the physical state of the disk. 
In particular, we shall examine whether the rate of magnetic flare generation 
within a radiation-dominated disk is diminished relative to that in a gas 
dominated configuration.  If this holds true, then the model would predict that 
radiation-dominated accretion disks should be in a softer state, while gas 
dominated disks should be in the hard state. Finally, we will compare the model 
predictions with the available observations of GBHCs and Neutron star (NS) systems.

\section{Energy Transport due to Magnetic Flares}

The reasons why strong X-ray emission might be expected from magnetic flares 
in accretion disks were first pointed out by Galeev et al. (1979), who showed that 
magnetic field dissipation inside the disk is not efficient in limiting the
magnetic field growth. The field is therefore likely to grow to its  
equipartition value and then be expelled from the disk due to magnetic buoyancy. 
Once outside, the field can in principle transfer its energy to particles by 
reconnection or some other mechanism, not unlike the manner with which the
Sun releases energy in flares (e.g., Parker 1979; Priest 1982). The work of 
Galeev et al. (1979) shows that magnetic flares may in this fashion account for 
a substantial energy outflux.  The fact that the differential rotation in
disks is stronger than in stars might also imply a stronger turbulence,
convection and therefore, relatively speaking, possibly even a more intense 
magnetic field than is seen in stellar environments.

At the same time, the standard accretion disk theory of Shakura \& Sunyaev (1973) 
invokes only radiation transport as the mechanism for transferring the gas 
thermal energy out of the disk (see Bisnovatyi-Kogan \& Blinnikov 1977). 
We shall try to account for the energy being transported out of the disk by
magnetic flares with a simple parameterization (cf. Svensson \& Zdziarski
1994; hereafter SZ94).  The vertical velocity associated with the flux tube
buoyancy is limited to the gas sound speed, and the magnetic pressure is limited 
by the gas plus radiation pressure.  Accordingly, we will parameterize 
the magnetic energy flux for the gas dominated region of the disk as 
$F_{\rm m} \equiv \beta c_s P$, where $\beta$ is a dimensionless parameter 
of order $1$ or less. In equilibrium, assuming that the inward advection of
energy in the disk is not dominant, the gravitational energy dissipation
rate should be approximately equal to the power carried outwards by the
radiation (with a flux $F_{\rm r} = c P_{\rm r}/\tau_{\rm d}$, in terms of
the radiation pressure $P_{\rm r}$ and the disk vertical optical depth
$\tau_{\rm d}$) and the magnetic field.

Vishniac (1995) has argued that
$\beta$ is likely to be of order $\alpha^2$, where $\alpha$ is the viscosity 
parameter in a standard accretion disk.  With this prescription, the
magnetic flare energy flux would only be a small ($\sim \alpha$) fraction 
of the radiative power from the disk. In a longer follow-up paper 
(Nayakshin \& Melia 1997e, hereafter paper II) we shall discuss why $\beta$ 
can be much larger than this estimate for magnetic flux tubes with a large
spatial scale (i.e., $\sim H$, where $H$ is the disk scale height). We point
out that $\beta$ may be as large as  $\alpha$, which
will be interpreted as an indication that the magnetic field energy transport 
can be extremely efficient compared to that due to radiation
(see discussion below and Eq. (1)).

Before investigating the consequences of this parameterization of the 
magnetic energy flux, let us first discuss the analogous situation in 
radiation dominated disks. Here, the physics is not as clear cut since 
the radiation does not interact directly with the magnetic field, but rather
via the particles. As is well known (e.g., Parker 1979), the maximum
pressure within a magnetic flux tube is set by the external pressure. However,
when the flux tube is optically thin, even if it were radiation free
initially, the photons diffusing into it would eventually raise the 
internal radiation pressure to the external value, and the internal
magnetic field pressure could then be only as large as the gas pressure. 
In paper II, we will provide an estimate showing that this is indeed the case
for accretion disks in the standard theory (see also Sakimoto \& 
Coroniti 1989; Stella \& Rosner 1985). We expect
that in the radiation-dominated regime, most of the magnetic field is diffuse, 
or at best that it forms weak flux tubes, and that it therefore results
in a more nearly uniform heating of the upper disk atmosphere as opposed to
localized high-energy magnetic flares. 

To formalize in the simplest way the foregoing discussion in our modeling
of the magnetic flare flux, $F_{\rm m}$, we will write $F_{\rm m} =
\beta c_{\rm g} P_{\rm g}$ for both the gas- and radiation-dominated
disks, in which $c_{\rm g} \equiv [P_{\rm g}/\rho]^{1/2}$, and $P_{\rm g}$ 
and $\rho$ are the gas pressure and density, respectively. This parameterization 
assumes that the radiation is completely decoupled from the magnetic field, so 
that only the gas pressure can contribute to the strength of the flare. 
A more sophisticated approach than this will be given in paper II. 
Here we use the simpler approach for clarity, 
since it provides a good overall description of the magnetic flux and is well
suited for comparing to observations of GBHCs. Using the standard disk equations 
(e.g., Frank, King \& Raine 1992), one can show that the total dissipation
rate of the gravitational energy per unit area of one side of the disk 
is $D(R) = (9/8) \alpha c_s P$. In a steady state disk with negligible
advection, $D(R)$ should be equal to the magnetic plus radiative
flux. Therefore, for the radiation flux one can write
\begin{equation}
{c P_{\rm rad}\over \tau_{\rm d}} = (9/8) \alpha c_s P 
\left [1 - {8\beta\over 9\alpha} \left({P_{\rm g}\over
P_{\rm g} + P_{\rm rad}}\right )^{3/2}\right ]\;.
\end{equation}
This expression is in the form of Equation (4) of SZ94, 
who assumed that only a fraction $1-f$ of the gravitational energy 
is dissipated inside the disk (and subsequently is transferred out by 
radiation), whereas the remaining fraction $f$ is transferred to the 
overlying optically thin corona and is dissipated there. We thus see
that the fraction $f$ of SZ94 can be interpreted in our picture to
be $(8\beta/9\alpha)\, P_{\rm g}^{3/2}/(P_{\rm g} + P_{\rm rad})^{3/2}$.
However, we caution that locally, the ratio $f_{\rm loc}$ of the X-ray flux 
from the AR to the local intrinsic disk flux {\it during the flare} is still much higher 
than $f/(1-f)$, because the latter is the ratio averaged over the whole
disk, i.e., it includes regions where flares are not active at a given time. 
To reconcile these two quantities, we introduce the covering fraction
$f_{\rm c}\ll 1$ of the magnetic flares (see Haardt et al. 1994), defined 
as the ratio of the total area covered by the active regions at a given 
time to the whole area of the disk enclosed by a characteristic size $R_g$ 
dominating the emission, such that $f_{\rm loc} = f_{\rm c}^{-1}\,f/(1-f)$. 
This allows the X-ray spectrum to be steep even in cases where $f\lesssim 1/2$.

\section{Spectral Transitions in Accretion Disk Systems}

We will now attempt to use our very simple model for the magnetic flux to
delineate the distinct regions of the solution phase space.  The most 
important parameter is the dimensionless accretion rate, $\dot m$, defined as
$\dot{m} \equiv \dot{M}c^2/L_{\rm Edd}$, where $\dot{M}$ is the actual 
accretion rate, and $L_{\rm Edd} = 1.26\times 10^{39} M_{10}$ erg/sec
is the Eddington
luminosity for a 10 Solar mass black hole ($M_{10}\equiv M/10 \msun$).
First of all, it is already clear that the model predicts that the
X-ray spectrum softens as the accretion rate increases and the radiation 
pressure starts to dominate over the gas pressure, since the importance 
of magnetic flares as the mechanism for generating an outward energy flux 
decreases. At the same time, the vertical optical depth of the accretion 
disk decreases (see SZ94) with increasing $\dot{m}$, and so the average 
photon finds it easier to escape from the disk, thus making radiation 
transport more important compared to the flares.

To quantify this situation, we shall seek an approximate description of
the ratio $x\equiv P_{\rm rad}/P_{\rm g}$ for all accretion rates. We do this by
taking the average of this ratio (found in SZ94) for the radiation-dominated
and gas dominated disks. Also, we define $f_0 \equiv (8\beta/9\alpha)$, which
is the limiting value of $f$ for $x\ll 1$. The fraction $f$ as a function of $x$
is then $f(x) = f_0 /(1 + x)^{3/2}$. Carrying this derivation out, we obtain an
implicit relation between $\dot{m}$ and $x$:
\begin{equation}
x = 1/2\, \left [ A \dot{m}^2 (1-f(x))^{9/4} + B \dot{m}^{4/5}
(1-f(x))^{9/10}\right ]\;,
\end{equation}
where $A\equiv 12.8\, (\alpha M_1)^{1/4} r_6^{-21/8} [J(r)/0.3]^2$, and
$B\equiv 3\, (\alpha M_1)^{1/10} r_6^{-21/20} [J(r)/0.3]^{4/5}$, with
$r_6 = r/6$, $J(r) = 1 - \sqrt{3/r}$ and $r$ is the distance from the 
black hole in terms of the gravitational radius.

In Figure 1, we show the run of $1-f(x)$ versus the accretion rate $\dot{m}$.
The curves are labeled by their corresponding value of $f_0$. For $1-f(x)\ll 1$,
the accretion disk solution is that dominated by magnetic flares. In the
context of our model, this region corresponds to the hard states of GBHCs 
and NS systems (see Nayakshin \& Melia 1997a for a discussion of the
Cygnus X-1 hard state spectrum). The region to the right of the ``kink'' 
in the curves corresponds to the soft states. However, we should keep in mind 
the fact that this very simple model yields the {\it minimum} $\dot{m}$ for 
which the transition can occur, since we have assumed that the radiation is 
completely decoupled from the magnetic field, so that the flares subside
as $P_{\rm rad}/P_{\rm g}$ becomes larger than 1.  At the same time, the 
accretion disk is still optically thick at $P_{\rm rad}/P_{\rm g}\sim 1$, as 
one may confirm using the results of SZ94 (assuming $\dot{m} \sim 1$). 
Therefore, we should also investigate the possibility that the transition 
occurs at a higher value of the ratio $P_{\rm rad}/P_{\rm g}$. To account
for this qualitatively, we assume that $f(x)$ may be written as $f(x) = f_0/
(1 + (x/x_c))^{3/2}$, where $x_c\geq 1$. The physical meaning of the parameter
$x_c$ is that the radiation decouples from the gas when $P_{\rm rad}/P_{\rm g}$
is larger than $x_c$. The run of $1-f(x)$ versus the accretion rate $\dot{m}$ 
is in this case shown in Figure 2, where the solid curve corresponds to the
case $x_c = 1$. 

Comparing Figures 1 and 2, we see that magnetic flares are important in the 
overall disk energy balance when the parameter $f_0$ is close to 1,
i.e., $1-f_0\ll 1$. If the actual value of $x_c$ is close to 1, then
the  transition from flare-dominated to radiation transfer dominated disks
is relatively gradual, and occurs over a dynamic range in $\dot{m}\gtrsim 3$. 
However, for $x_c\sim 3$, the transition can be very abrupt, and may be triggered 
with just a $30\%$ change in $\dot{m}$. This is interesting in view of 
the fact that recent state transitions in Cygnus X-1 have been observed to
occur with at most a $30\%$ change in the overall disk luminosity (proportional to 
$\dot{m}$ in this model) during a hard-soft-hard cycle (Zhang et al. 1997b). 
Finally, $x_c\gtrsim 3$ may lead to two values of $f$ for the
same $\dot{m}$, i.e., the disk may be either in the hard or the soft state. It
is not yet clear whether the hard state is stable against decay to the soft
state in this region of phase space, since a vertical perturbation
of the disk may result in a decoupling of the radiation from the flares and a 
consequent transition from the hard to the soft state. Moreover, with our simple 
parameterization, it is difficult to judge what range of $x_c$ is physical.
This question will be pursued in the future, but for now we will
study cases with $x_c=1$ and $2$ as representative values in order to 
compare the predictions of our model with the observations.

\section{Phase Space Trajectory of Transient GBHCs and NS Systems}

Recently, Barret, McClintock \& Grindlay (1996) assembled a sample of
GBHCs and several other transient sources in a $L_{\rm hx}-L_{\rm x}$ phase space,
where $L_{\rm hx}$ is the hard X-ray luminosity in the range $20-200$ keV, and
$L_{\rm x}$ is the X-ray luminosity in the range $1-20$ keV. One of
the striking results of this exercise is that both neutron stars and GBHCs always
have relatively soft spectra when they radiate at a high fraction of
their Eddington luminosity. In particular, none of the former show a 
hard tail in their spectrum at a luminosity higher than about $2\times 
10^{37}$ erg s$^{-1}$, or $\sim 0.1$ of the Eddington luminosity for a 
$1.4\;\msun$ NS. For GBHCs, this dividing luminosity is a fraction of
$10^{38}$ erg s$^{-1}$, or about $0.03 - 0.1\times L_{\rm Edd}$ for a
$5-10$ $\msun$ black hole. Similarly, Grove, Kroeger \& Strickman (1997) 
and Grove et. al. (1998) showed that GBHCs occupy at least four spectral
states in order of decreasing X-ray luminosity. In particular, they found that
when the X-ray luminosity (above $1$ keV) is at about the Eddington limit, the
spectrum is dominated by the so-called ultrasoft blackbody component with
$kT\sim 1$ keV; a weak hard tail is seen above $\sim 10$ keV, and rapid intensity
variations are present. At lower luminosities (typically $\sim 0.1 L_{\rm Edd}$),
the spectrum again shows an ultrasoft component and a weak hard tail, but 
rapid intensity variations are weak or absent. The hard state exhibits a
single power-law spectrum with a photon number index $\Gamma\sim 1.5-2$, and
corresponds to a luminosity in the range $10^{36-37.5}$ erg s$^{-1}$. Finally, 
for the low luminosity, quiescent state, $L_{\rm X} < 10^{34}$ erg s$^{-1}$, 
but its spectral shape is not very well known. We note that the luminosity
above 1 keV should be close to the disk's bolometric luminosity
since the temperature in the soft state is about $1$ keV, and thus 
most of the blackbody power should lie above $1$ keV, whereas in the hard state
the intrinsic disk temperature can be much smaller, i.e., $\gtrsim 0.1$
keV, but its contribution to the overall spectrum is negligible compared
to that of the hard power-law (see, e.g., Gierlinski et al. 1996).

Interpreting these intriguing results in terms of the accretion rate 
(assuming a rest mass energy conversion efficiency of 0.06), we see that 
the accretion rate dividing the hard and soft states seems to be 
$\dot{m}_{\rm crit}\sim 1$ for GBHCs and $\sim 1.6$ for NS. Positioning 
these values in Figure 2, it is evident that those curves with $x_c = 
P_{\rm rad}/P_{\rm g} \sim$ $1-$ few correspond to a transition from the 
flare-dominated to the no-flare regimes at $\dot{m}\sim 0.3 - 2$. 
In addition, we note that the intrinsic universality of the two-phase model
is well suited to explain the relatively broad luminosity range ($10^{36-37.5}$ 
erg s$^{-1}$) covered by the hard state. Indeed, this spectral universality 
was the primary reason for the original application by Haardt \& Maraschi (1991)
of the two-phase model to Seyfert Galaxies, whose spectra display
the ``standard'' intrinsic X-ray photon index $\Gamma\sim 1.9$. 
Namely, once $1 - f(x)\ll 1$, the X-ray spectral slope and the ratio 
$L_{\rm hx}/L_{\rm x}$ are dictated by the reflection/reprocessing of
X-rays at the surface of the underlying cold disk.
It is unlikely that these processes will scale significantly 
with the bolometric luminosity in GBHCs (see, e.g., Nayakshin \& Melia 
1997a,b), and we therefore expect for the region $1 - f(x)\ll 1$,
that the ratio $L_{\rm hx}/L_{\rm x}$ will be approximately independent 
of the source luminosity. However, this situation is bound to break down at 
some relatively low luminosity, since one of the key
assumptions of the model is that the formation of the spectrum is due
to the Compton up-scattering of the lower energy photons. At the same time,
the compactness of the AR is decreasing with decreasing $\dot{m}$,
and at $l\lesssim 0.5$ bremsstrahlung takes over as the leading
radiation mechanism (see Fabian 1994). In addition, the explanation for 
the observed optical depth of Seyferts and GBHCs proposed by Nayakshin \& 
Melia (1997c) requires the compactness to be larger than unity as well. 
Using the estimates of Nayakshin \& Melia (1997d), it is possible to show 
that the lowest luminosity for which the two-phase model is still valid 
is likely to be of order $10^{-3}$ $\times L_{\rm Edd}$ (using their 
Eq.[5] with $\alpha\sim 0.1$, $b\sim 10$, and with a magnetic pressure within
the flux tube set at the disk equipartition value; a smaller magnetic 
pressure will only increase this luminosity). Below this minimum luminosity, 
the spectrum should be quite different from that in the hard state, and it
is likely to be dominated by the soft intrinsic disk emission. A power-law 
tail similar to the Solar X-ray spectrum may be present as well, since a
compactness $\ll 1$ is also pertinent to the magnetic flares in the Sun.
Accordingly, both the minimum and maximum luminosities of the hard state
in GBHCs with $M\sim 6 \msun$ seem to be consistent with the predictions
of the magnetic flare-disk model. 

Barret et al. (1996) also plotted the evolutionary track of the GBHC
transient source GRS 1124-68 on the same $L_{\rm hx}-L_{\rm x}$ 
phase diagram. During its evolution, this source occupied both a hard 
and a soft state, and it spanned a broad range in luminosity. It therefore 
constitutes an especially interesting object to study in the context of
our model. To simulate its spectral evolution, we assume the spectral model 
of Nayakshin \& Melia (1997a) in order to describe the hard spectral state 
of the accretion disk. For a given $\dot{m}$, we can find the corresponding 
value of $x$ from Equation (1), which allows us to compute $f(x)$, i.e., 
the fraction of the gravitational energy being deposited within the active
region. For simplicity, we also assume a fixed covering fraction $f_c = 0.2$; 
our results depend very weakly on this value since the phase space trajectory 
is decided mostly by the evolution of $f(x)$, describing the partitioning of the 
total available power into the cold disk and flare components. 
It is then possible to determine
the spectrum for a broad range of $\dot{m}$, and in Figure 3, we show the
decomposition of this spectrum into the hard ($L_{\rm hx}$) and soft 
($L_{\rm x}$) components (for two values of $x_c$---see the caption to the 
Figure), together with the data for GRS 1124-68  
(adopted from Barret et al. 1996). This figure also includes the data
for other BHBs (i.e., for GBHCs that have well established mass estimates),
whose $L_{\rm hx}$ and $L_{\rm x}$ are inferred from Figure 1 of Barret et 
al. (1996), using their Table 1 for the black hole masses. The Cygnus X-1 
data are also plotted for reference. Within the framework of this model, 
the hard X-rays in both the hard and soft states of Cygnus X-1 are 
created by the same mechanism, viz. magnetic flares, but in the soft state 
most of the total accretion power is dissipated inside the accretion 
disk, which acts to soften the spectrum.  Here, the disk's effective 
temperature $T_{\rm bb}$ is assumed to evolve such that $F_{\rm s} = 
\sigma_B T_{\rm bb}^4$, where $F_{\rm s}$ is the intrinsic disk flux and 
$\sigma_B$ is the Stefan-Boltzmann constant.

It is evident from Figure 3 that the lower part of the phase space
trajectory, within the region where $L_{\rm x}\lesssim 0.2$, is described quite 
well by the magnetic flare model. As discussed above, the region with 
$L_{\rm x}\lesssim 0.02$ corresponds to parameter values where
the flares are efficient in producing hard X-rays, and the model predicts
a more or less constant ratio $L_{\rm hx}/L_{\rm x}$. Above 
$L_{\rm x}\lesssim 0.02$, the hard X-ray luminosity declines sharply 
due to a decrease in the efficiency of magnetic flare generation, 
since the radiation pressure, constituting the largest reservoir of
available energy when $P_{\rm rad}\gg P_{\rm g}$, decouples from the
magnetic field.

What happens at a still higher value of $\dot{m}$ where the model cannot 
reproduce the sharp rise in the hard X-ray luminosity? This question is 
beyond the scope of the present paper, since this high accretion rate
is clearly outside the region of applicability of the magnetic flare model. 
For such high values of $\dot{m}$, the magnetic flares become ineffective 
in transporting the dissipated gravitational energy out of the disk. 
At the same time, it is well known that radiation dominated disks are 
unstable, and thus it may well be that the disk finds another mechanism for 
transporting energy outwards. For example, the disk may establish a full 
corona, heated by hydromagnetic waves, or the disk may evolve into the 
two-temperature configuration of Shapiro, Eardley \& Lightman (1974; 
see also Bisnovatyi-Kogan \& Blinnikov 1977). Whatever its nature, the disk 
must clearly use a different mechanism for producing the X-rays because the 
dip in the $L_{\rm hx}-L_{\rm x}$ plane is very sharp, suggesting that
one emissivity law is being phased out while a second is being turned on. 
We will address this question more fully in future work. There also remains
a great deal to be done with Neutron star systems, which we have only
touched on here.  Nevertheless, the qualitative luminosity-spectral 
state relationship for NSs should be similar to that of the black 
holes, and on this basis we conclude that the spectral evolution of GX-339 
shown in Figure 3 of Barret et al. (1996) identifies it as either a black 
hole positioned further away ($\sim 6$ kpc) than usually assumed, or 
more likely, as a neutron star.

\section{Discussion}

We have shown that the magnetic flare-disk model makes clear predictions 
about the relationship between the spectrum and the accretion rate (or
equivalnetly, the total disk luminosity). In particular, this model predicts 
that more luminous sources should have softer X-ray spectra, if their 
luminosity is larger than about 5\% of their Eddington limit. 
We have compared the model predictions with the available data for
GBHCs and NS systems, and have found a good agreement with the observations
within the uncertainties.  Perhaps the most important conclusion we
can make is that if the luminosity-spectral state relationship
reported here is confirmed by future studies, this representation 
may allow us to estimate the mass of disk-accreting X-ray sources whose
bolometric luminosity is known, and may thus prove to be a valuable
tool in both galactic and extragalactic X-ray astronomy.

\section{Acknowledgments}

This work was partially supported by NASA grant NAG 5-3075.

%
%
%

{}

\vfill\eject
\figcaption []{Fraction of the energy outflux due to radiation
transport as a
function of the dimensionless accretion rate. Curves are labeled by their
corresponding values of $f_0$ (see text). Notice that in the region with
small $\dot{m}$, the energy outflux is mostly due to flares, whereas for
large $\dot{m}$ the radiation transport is most efficient.}

\figcaption []{Same as Figure 1, for $f_0 = 0.99$ and for three
different values of the parameter $x_c$, i.e. the ratio of $P_{\rm
rad}/P_{\rm g}$ where magnetic flares become inefficient in the energy
transfer out of the disk (see test).}

\figcaption []{Space phase trajectory of an accretion disk system 
in the hard-soft X-ray luminosity coordinates for $x_c = 1$, $\eta = 0.12$
(solid curve) and $x_c = 2$, $\eta = 0.06$ (dashed curve), where $\eta$ is the
efficiency of converting gravitational energy into radiation. 
Data for the transient
source GRS 1124-68 evolution are shown with the dots connected by the
dashed line. Cyg X-1 1970's hard and soft states 
(see explanation in Barret et al. 1997) 
are shown with large polygons, whereas newer data due to Zhang
et. al. (1997b) are shown by the two square boxes. Filled triangles
represent the other well established black hole systems from Fig. 1 of
Barret et. al. (1997).}

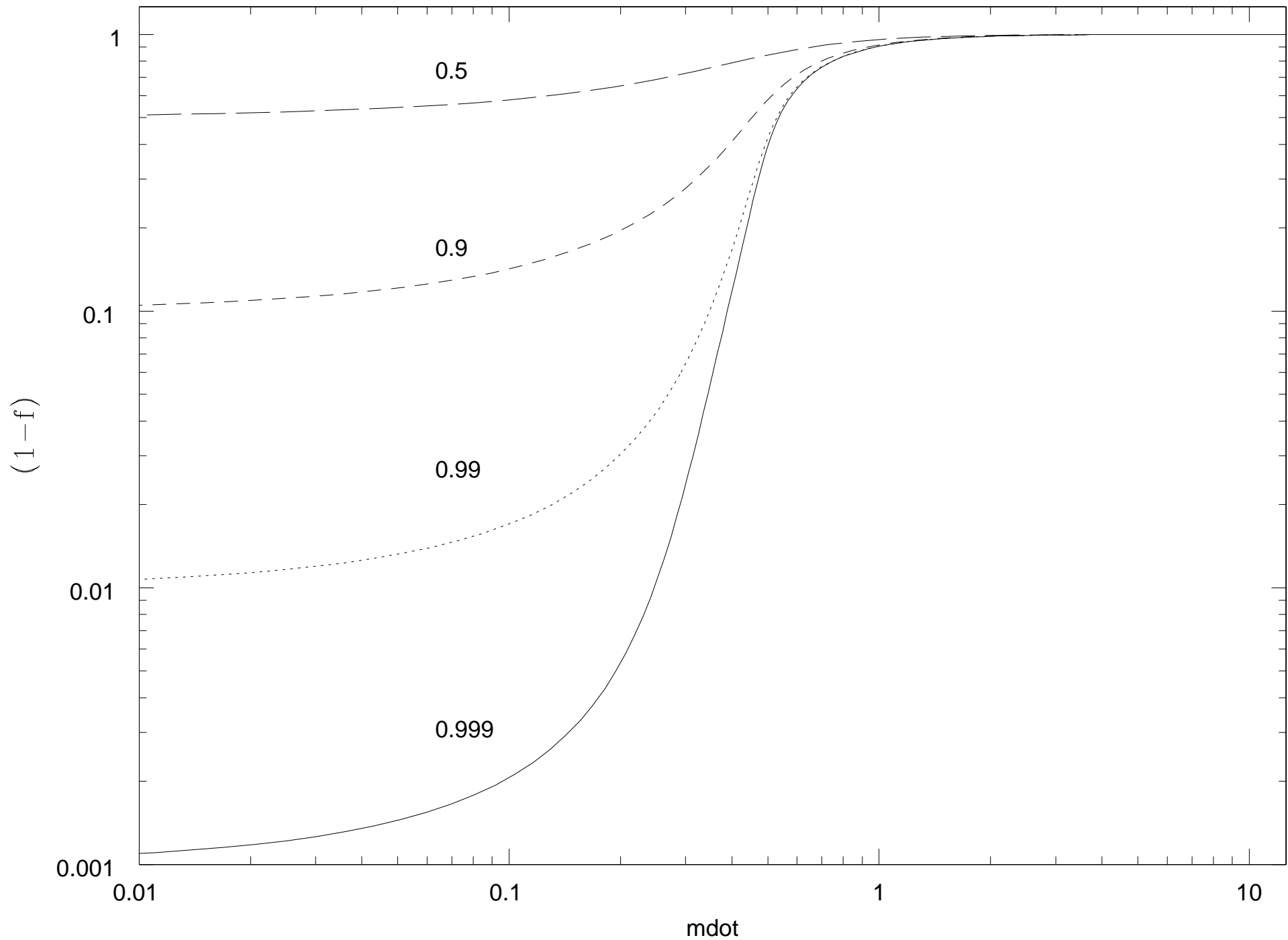
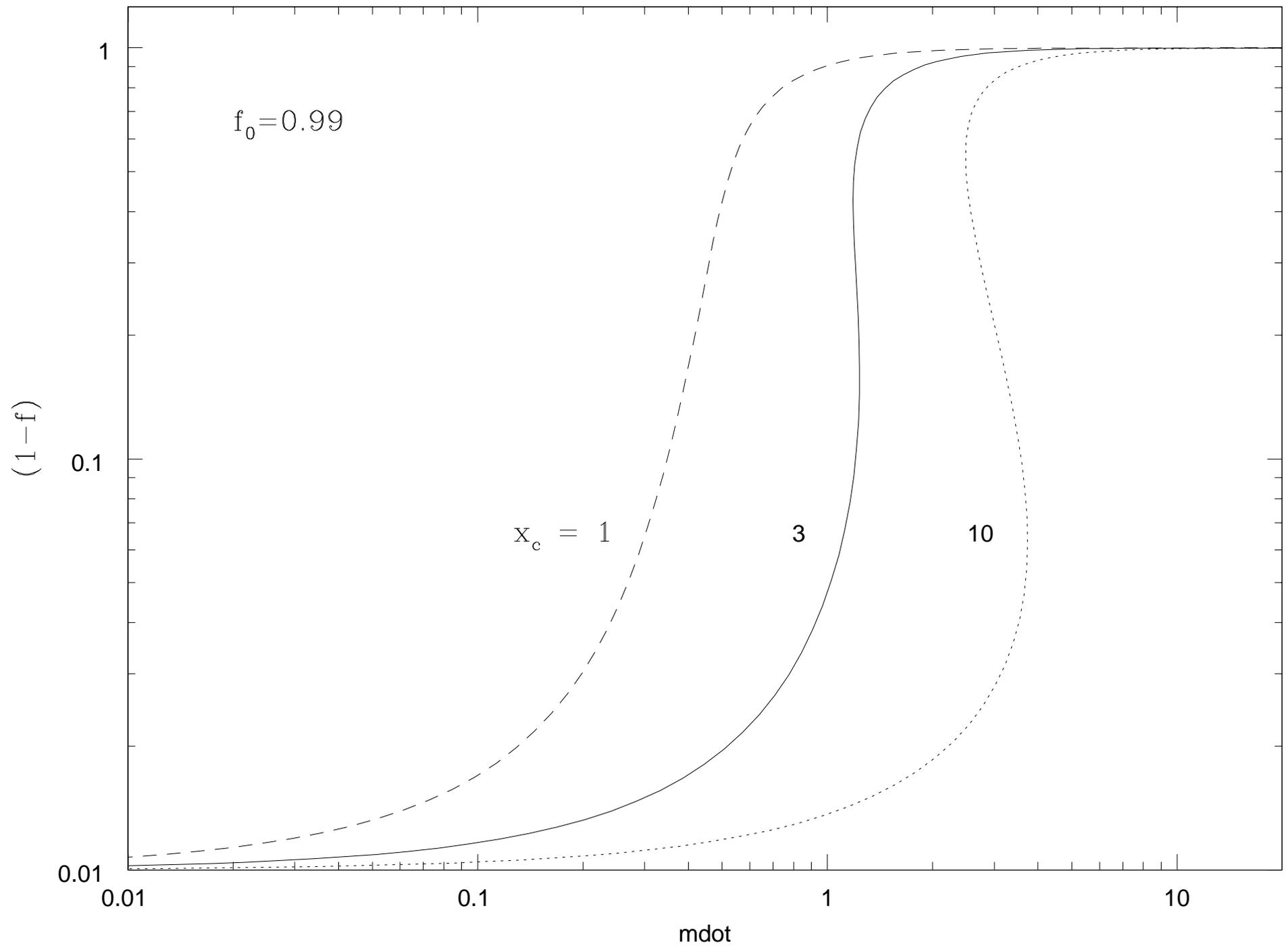
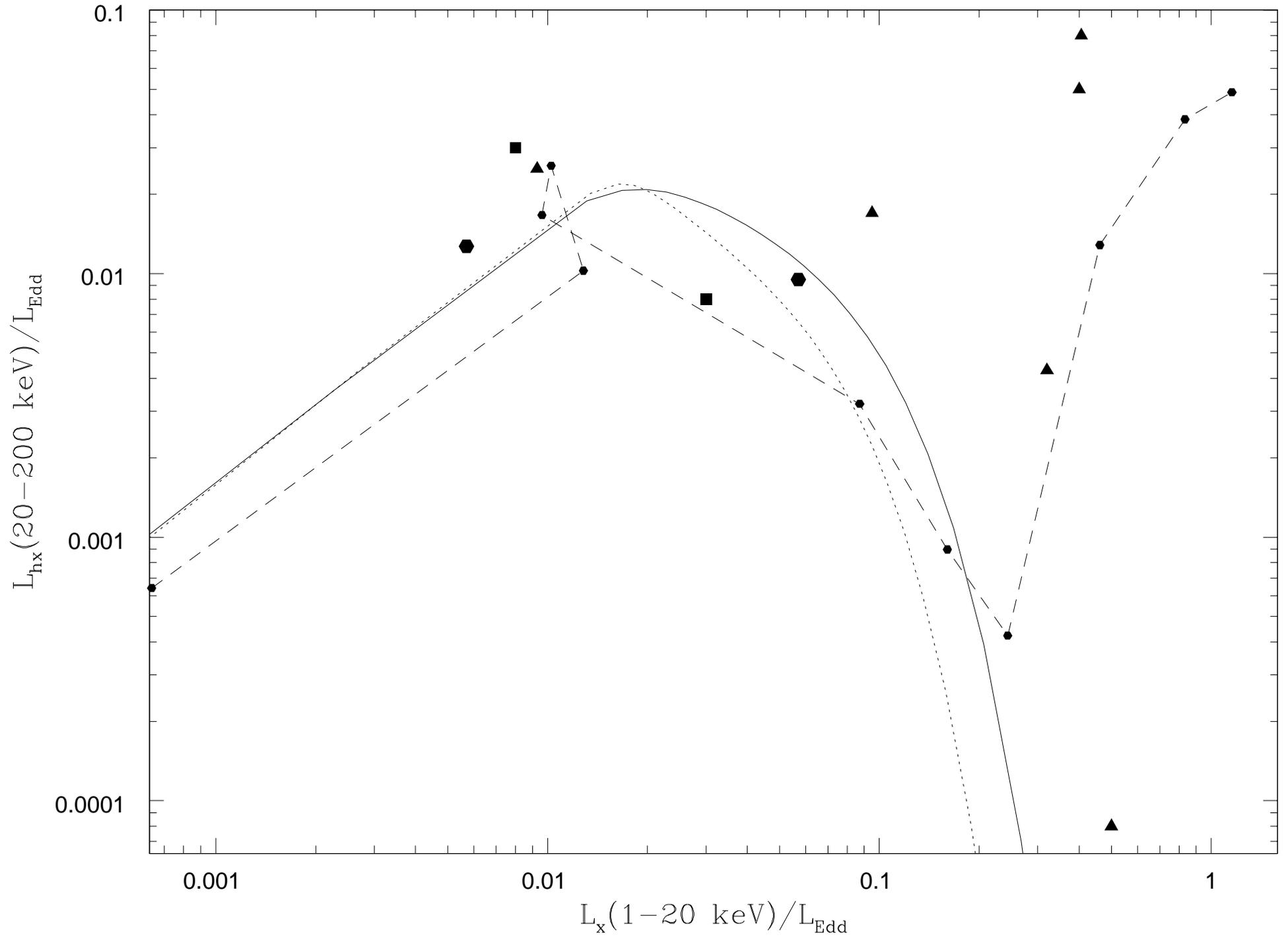
\end{document}